%% file: main.tex
\documentclass[10pt,conference]{IEEEtran}

\usepackage[utf8]{inputenc}
\usepackage{algorithm}
\usepackage[noend]{algpseudocode}
\usepackage{amsmath}
\usepackage{graphicx}
\usepackage{xcolor}
\usepackage[inline]{enumitem}
\usepackage{smartdiagram}
\usepackage{comment}
\usepackage{todonotes}
\usepackage{soul}
\usepackage{float}
\usepackage{subfigure}
\usepackage{comment}

\usepackage[url=false,doi=false,isbn=false,eprint=false,firstinits=true,maxnames=10]{biblatex}
\input{macros}

\begin{document}
\title{Scalable Statistical Root Cause Analysis on App Telemetry}

\author{
\IEEEauthorblockN{Vijayaraghavan Murali}
\IEEEauthorblockA{Facebook, Inc.\\
U.S.A. \\
vijaymurali@fb.com}
\and
\IEEEauthorblockN{Edward Yao}
\IEEEauthorblockA{Facebook, Inc.\\
U.S.A. \\
edwardcdy@fb.com}
\and
\IEEEauthorblockN{Umang Mathur}
\IEEEauthorblockA{Facebook, Inc.\\
U.S.A. \\
umathur@fb.com}
\and
\IEEEauthorblockN{Satish Chandra}
\IEEEauthorblockA{Facebook, Inc.\\
U.S.A. \\
schandra@acm.org}
}

\maketitle

\date{August 2020}

\input{abstract}

\input{intro}

\input{overview}

\input{rca-framework}

\input{rca-practice}

\input{rca-facebook}

\input{evaluation}

\input{related}

\input{conclusions}

\printbibliography

\end{document}

%% file: macros.tex
\newtheorem{definition}{Definition}
\newtheorem{lemma}{Lemma}

\newcommand{\figlabel}[1]{\label{fig:#1}}
\newcommand{\figref}[1]{Fig.~\ref{fig:#1}}

\newcommand{\edward}[1]{{\color{green} [~\emph{EY:~#1}~]}}

\newcommand{\revision}[1]{#1}
\newcommand{\ignore}[1]{}

\newcommand{\name}{Minesweeper}
\newcommand{\fone}{F_1}

\newcommand{\emptyseq}{\varepsilon}
\newcommand{\minsup}{\mathtt{min\_support}}

\DeclareNameAlias{sortname}{last-first}

%% file: abstract.tex

\begin{abstract}

Despite engineering workflows that aim to prevent buggy code from being deployed, bugs still make their way into the Facebook app.
When {\em symptoms} of these bugs, such as user submitted reports and automatically captured crashes, are reported, finding their {\em root causes} is an important step in resolving them.
However, at Facebook's scale of billions of users, a single bug can manifest as several different symptoms according to the various user and execution environments in which the software is deployed.
Root cause analysis (RCA) therefore requires tedious manual investigation and domain expertise to extract out common patterns that are observed in groups of reports and use them for debugging.

We propose \name{}, a technique for RCA that moves towards automatically identifying the root cause of bugs from their symptoms.
The method is based on two key aspects: (i) a scalable algorithm to efficiently mine patterns from telemetric information that is collected along with the reports, and (ii) statistical notions of precision and recall of patterns that help point towards root causes.
We evaluate \name{}'s scalability and effectiveness in finding root causes from symptoms on real world bug and crash reports from Facebook's apps.
Our evaluation demonstrates that \name{} can perform RCA for tens of thousands of reports in less than 3 minutes, and is more than 85\% accurate in identifying the root cause of regressions.

\end{abstract}

%% file: intro.tex
\section{Introduction}
\label{sec:inro}

    

When code is shipped at Facebook, it goes through the general software engineering processes of code review, testing, and static analysis.
Despite this, bugs get inadvertently shipped out into production.
These bugs end up causing either crashes in the field or functional issues in product usage.
Crashes are automatically captured and reported back, and users manually submit bug reports if they encounter any functional issues.

To maintain app quality, crashes and bug reports need to be fixed as soon as possible, 
and quick identification and isolation of their root causes is the first point of attack.
In the presence of high-signal debugging information such as stack traces,
localization techniques like Scaffle~\cite{pradel2020scaffle} can be useful.
However, for a large portion of bug reports, stack traces are unavailable.
For instance, a crash due to device running out of memory (OOM) would not contain a stack trace, as there would not be enough memory to capture it.
A user-submitted bug report due to functional issues such as a failed photo upload, would not even have a crash to begin with, and would only contain a simple non-technical bug description.

Root cause analysis (RCA) of these bugs is extremely difficult due to the lack of debugging signals at the time of manifestation of the bug.
In such cases, engineers have to rely on telemetry that captures other properties of the bug, such as device features or events that precede the bug.
In these contexts, RCA involves aggregating multiple reports and investigating patterns in telemetry that are distinctive to one group of reports compared to other groups.
For instance, when a user submits a bug report, a free-form bug description such as ``photo upload failed'' could be useful for aggregating multiple reports of the bug, but is not likely indicative of the root cause of the bug.
Often, in addition to the bug description, the error reporting system would have captured a sequence of actions the user performed in the app prior to encountering the bug.
Software engineers can then look for patterns in those sequences that are distinctive to users who encountered the bug compared to those who did not.
Once a sequence of actions highly correlated with the occurrence of the bug is identified, it can help in the task of attributing the bug to the right developers and possibly even in reproducing the bug. 

However, this process is largely manual, requiring dedicated human effort to analyze bug reports and determine what, if any, are the observable patterns distinctive to some crash or bug report.
As a result, root cause analysis is typically slow and cumbersome, and requires domain expertise to understand telemetry associated with bug reports.
Unsurprisingly, this could lead to either inaction on the bug reports, or potentially chasing spurious signals and wasting time. 

In this paper, we propose \name{}, a technique for RCA that expedites and lowers the domain knowledge barrier for debugging errors.
\name{}-based RCA is completely automated, scalable, and is based on formal statistical concepts. 
A high-level overview of \name{} is as follows.
We begin with the notion of an {\em event}, which records some telemetric information 
about the app and its environment, such as the surface the user is in the app, or available device memory,
captured at a point in time.
A {\em trace} is a sequence of events in chronological order, capturing telemetry as the user is using the app.
\figref{blackbox-trace} shows an example of a trace
generated when a user interacts with the Facebook app.
The different interactions in the user session corresponding to different events in the trace,
lined up according to chronological order.
A trace could either end in the user encountering a bug or crash, 
or terminate normally when, for example, the user closes the app.
A large amount of such traces are collected to be analyzed.
The core problem \name{} solves can then be formalized as follows:
given two groups of traces, one in which the traces contain the bug (the {\em test group}) 
and one in which the traces do not contain the bug (the {\em control group}), 
find patterns of events that are statistically distinctive to the test group as opposed to the control group.
Such patterns are likely to reveal insights into events that are correlated to the crash or bug, thereby pointing towards its root cause.

\begin{figure}
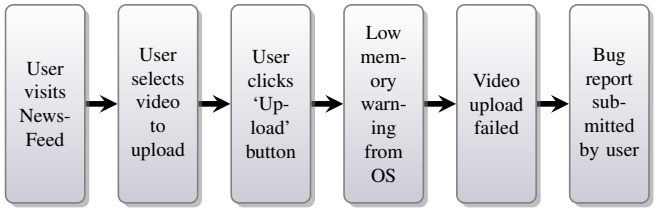

\centering
\scalebox{.75}{
\smartdiagramset{
uniform color list=blue!20!black!20!white for 6 items,
uniform arrow color=true,
arrow color=black,
font=\normalfont,
back arrow disabled=true,
module minimum width=40,
module minimum height=100,
text width=30,
module y sep=1.2,
module x sep=2
}
\smartdiagram[flow diagram:horizontal]{
User visits NewsFeed,
User selects video to upload, 
User clicks `Upload' button,
Low memory warning from OS,
Video upload failed,
Bug report submitted by user
}
}
\caption{Example of a chronological trace of events, ending in a bug report.}
\figlabel{blackbox-trace}
\end{figure}

There are several technical challenges that arise when designing a framework 
like \name{} meant to operate at the scale and ecosystem of Facebook.
First, in order to extract patterns of interest that point towards the root cause, we need to decide how exactly our patterns look like.
Second, we need an efficient algorithm to mine these patterns from a large volume of traces, typically tens of thousands in Facebook's setting.
Third, when we mine a large number of patterns from our traces,
many patterns are likely to emerge, and not all of these patterns may
be indicative of a bug.
To cater for this, we need to cast the RCA problem into a well-formulated statistical setting 
so that the patterns we do extract conform to human intuitions
and can be effective in isolating the real root cause.
Finally, we need to address practical 
challenges that include handling numeric data in traces, 
and other human-centric challenges such as avoiding showing redundant patterns to developers 
that could otherwise multiply developer effort.

Towards addressing these challenges, we make the following contributions in this paper:
\begin{itemize}
    \item We propose \name{}, an automated root cause analysis method that is based on extracting patterns from traces of telemetric information.
    
    \item We utilize the notion of {\em sequential patterns}~\cite{Agrawal1993} from the data mining community for performing RCA. We leverage the {\em PrefixSpan} algorithm~\cite{Pei2004} that is well-known for being highly efficient at mining sequential patterns from traces. We also propose a statistical method for ranking patterns that is effective at extracting the most distinctive patterns for RCA.


    \item We discuss and propose solutions for practical challenges that arise in an industrial setting like Facebook, such as
    \begin{enumerate*}[label=(\roman*)] 
    \item avoiding ``redundant'' patterns that are similar in explaining the root cause, and
    \item handling numeric data in traces.
\end{enumerate*}
    
    \item We evaluate \name{} on real world crash and bug reports from Facebook's mobile apps, and show that it can efficiently perform RCA of 10,000 reports in less than 3 minutes.
    We also show using real data that the root cause insights it reports are accurate more than 85\% of the time.
    We also discuss case studies of its usage at Facebook.

\end{itemize}

\ignore{
The rest of this paper is organized as follows.
Section~\ref{sec:overview} gives an overview of our method with an example.
Section~\ref{sec:technical} presents our main technical RCA model.
Section~\ref{sec:rca-fb} discusses how RCA is done at Facebook and some practical challenges that arise in a large industrial setting.
Section~\ref{sec:evaluation} presents implementation details of \name{} and results on its experimental evaluation.
Section~\ref{sec:related} discusses related work.
Finally, Section~\ref{sec:conclusion} concludes the paper.
}

%% file: overview.tex

\section{Overview \& Example}
\label{sec:overview}

In this section, we provide an overview of our RCA method and illustrate it with an example.

\subsection{Overview}
\label{subsec:overview}

Users of Facebook's apps often report encountering a problem with the product as the result of a bug in the app's code.
However, the description of the problem reported by the users alone is often not likely to be detailed enough for an engineer to debug the issue.
Thankfully, in addition to such a bug description, apps can record a trace, or a sequence of events that happened before the user encounters the problem.
Examples of events include visiting a particular surface of the app, or a button click, or even a low-memory warning from the OS.
The trace can then serve as a ``flight recorder'' containing useful information about events that preceeded a bug occurrence.
Particularly, large amounts of such traces can be analyzed to extract common patterns that are associated with the occurrence of the bug, which can point towards its root cause.
Such an analysis constitutes what we call statistical root cause analysis. 

\begin{figure}[t]
    \centering
    \includegraphics[width=\columnwidth]{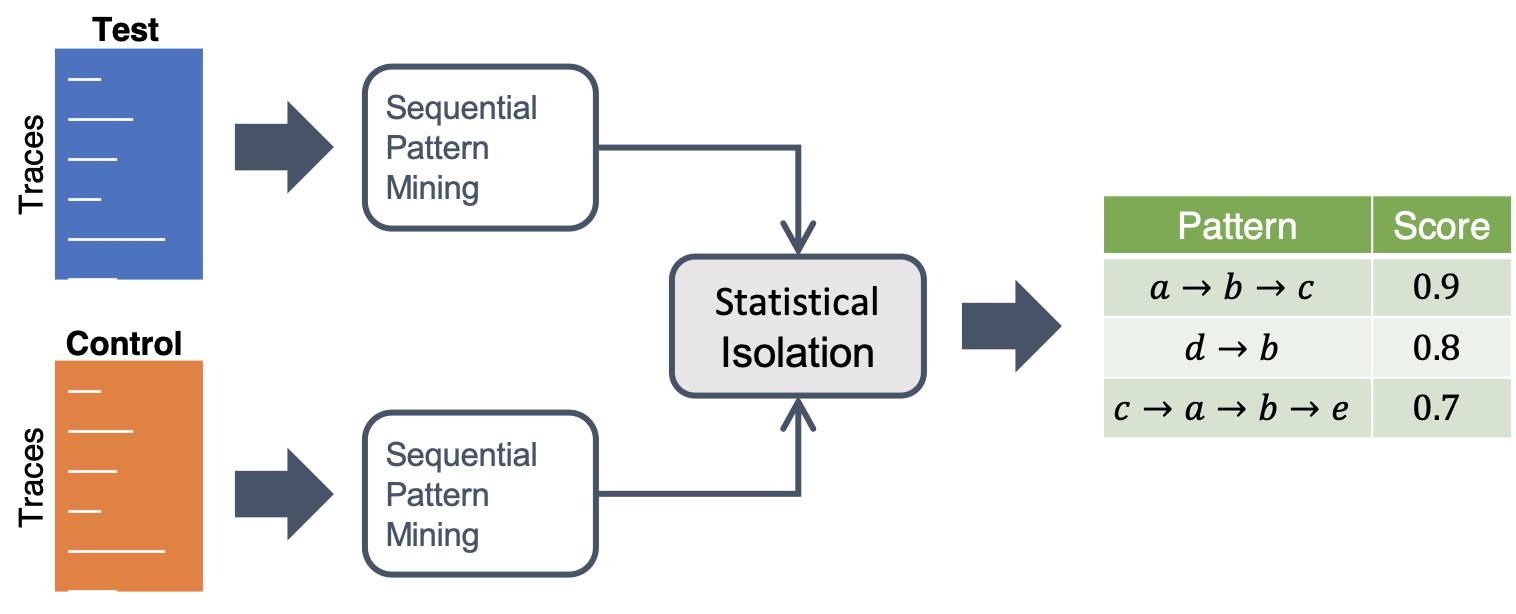}
    \caption{Overview of \name{}}
    \figlabel{overview}
\end{figure}

There are two main challenges in this process.
First, there is likely to be noise in the traces due to different users arriving at the bug following different events on the app.
This is due to the app's code being complex enough to have various control flows to reach a buggy program point.
Secondly, a pattern that is common among the users facing the bug does not necessarily imply that it is distinctive to the bug -- it could simply be a common pattern of usage.
\name{} handles these challenges using a two phase process of {\em pattern mining} and {\em statistical isolation}, as shown in~\figref{overview}.

Given a group of traces of events from users who had encountered the bug, \name{} begins by first extracting common patterns from the traces.
Specifically, it looks for {\em sequential patterns}, i.e., 
subsequences of 
events in the set of traces (Section~\ref{subsec:language}).
As an example, the sequential pattern $(e_1, e_2, e_3)$ matches a trace $\tau$ if 
$\tau$ has an occurence of the event $e_1$, followed by the event $e_2$, which is followed by
$e_3$, with possibly other events in between.
\name{} utilizes PrefixSpan, an efficient data mining algorithm (Section~\ref{subsec:scalable}) that can extract the most frequent patterns from a group of traces.

However, as mentioned, the most frequent patterns might not necessarily be distinctive to any bug or indicative of the root cause of any bug.
To address this, \name{} also takes as input a group of traces where the bug in question was {\em not} encountered.
In statistics terminology, the former group (buggy traces) is called the {\em test group}, and the latter {\em control group}.
As with the test group, \name{} also extracts patterns from the control group.
It then subjects all the patterns to statistical isolation.
The goal of this step is to rank patterns by their ``distinctiveness'' to the test group.

Distinctiveness is defined on the basis of two desired properties of a pattern -- (i) how prevalent in the test group this pattern is, and (ii) how unique to the test group (as opposed to the control group) this pattern is.
In information retrieval terminology, these are called {\em recall} and {\em precision}, respectively (Section~\ref{subsec:statistical}).
Ideally, one would like to maximize both quantities, but there is usually a trade-off between the two.
Hence, precision and recall are typically combined into a single score, called the {\em $\fone$ score}, using a harmonic mean.
The pattern with a higher $\fone$ score is more distinctive to the test group, and to the bug.
The final output of \name{} is list of sequential patterns in decreasing order of $\fone$ scores.

\subsection{Example}
\label{subsec:example}

\begin{figure}
{
\small
\begin{tabular}{cc}

Events: \{$e_1$, $e_2$, \ldots, $e_8$\}

\\ \\

{\bf Test group} $T$ & {\bf Control group} $C$ \\

\begin{tabular}{cl}
    $t_1$: & ($e_1$, $e_2$, $e_3$, $e_4$) \\
    $t_2$: & ($e_1$, $e_2$, $e_3$) \\
    $t_3$: & ($e_2$, $e_3$) \\
    $t_4$: & ($e_5$, $e_6$, $e_7$, $e_8$) \\
    $t_5$: & ($e_5$, $e_7$)
\end{tabular}

&

\begin{tabular}{cl}
    $t_6$: & ($e_1$, $e_2$, $e_4$) \\
    $t_7$: & ($e_1$, $e_3$, $e_4$) \\
    $t_8$: & ($e_1$, $e_3$) \\
    $t_9$: & ($e_6$, $e_7$) \\
    $t_{10}$: & ($e_5$, $e_6$, $e_8$)
\end{tabular}

\\ \\

\multicolumn{2}{c}{Top-5 \name{} patterns (without redundancy mitigation)} \\
\multicolumn{2}{c}{
\begin{tabular}{|c|c|c|c|c|c|}
\hline
{\bf Pattern} & \multicolumn{2}{c|}{\bf Support} & {\bf Prec-} & {\bf Recall} & {\bf $\fone$} \\ \cline{2-3}
& {\bf Test} & {\bf Control} & {\bf ision} && {\bf Score} \\
\hline
    $(e_2, e_3)$ & 3 & 0 & 1.0 & 0.6 & 0.75 \\
    $(e_2)$ & 3 & 1 & 0.75 & 0.6 & 0.67 \\
    $(e_3)$ & 3 & 2 & 0.6 & 0.6 & 0.6 \\
    $(e_1, e_2, e_3)$ & 2 & 0 & 1.0 & 0.4 & 0.57 \\
    $(e_5, e_7)$ & 2 & 0 & 1.0 & 0.4 & 0.57 \\
\hline
\end{tabular}
}

\end{tabular}
}
\caption{Example of \name's approach to extracting patterns for RCA}
\figlabel{example}
\end{figure}

Let us walk through an example with \name{}, using data depicted in \figref{example}.
Suppose that our app has 10 users, among whom 5 users reported a problem with its usage.
Also suppose  that there are 8 possible events tracked in the app: $e_1$ through $e_8$.
We have 10 traces in total on these events, 5 in the test group $T$ from users who encountered the bug, and 5 in the control group $C$ from the
remaining users.

\name{} then proceeds by extracting sequential patterns in $T$ and $C$.
For each pattern, it computes the {\em support} of the pattern -- the number of traces in which it appears -- in $T$ and $C$.
An example of a pattern that \name{} can extract here is $(e_1, e_3)$.
A trace matches this pattern if both $e_1$ and $e_3$ occur in it, in the same order.
This pattern matches 2 traces in $T$, and so its support in $T$ is 2.
Likewise, its support in $C$ is 2.
Similarly, the support of the pattern $(e_1, e_3, e_4)$ is $1$ in both $T$ and $C$.
As the space of patterns is combinatorial in nature, it is crucial to employ algorithms that 
navigate this space efficiently without an exponential blowup.

Once all patterns are extracted along with their supports in $T$ and $C$, \name{} performs statistical isolation.
For each pattern $P$, it computes precision and recall using its support:
\begin{align*}
\begin{array}{rcl}
\text{Precision}(P) &=& \dfrac{\text{support of } P \text{ in } T}{\text{support of } P \text{ in } T + \text{support of } P \text{ in } C} \\\\
\text{Recall}(P) &=& \dfrac{\text{support of } P \text{ in } T}{\text{total number of traces in } T}
\end{array}
\end{align*}

Informally, precision describes how accurate $P$ is in detecting if a given trace is in the test group rather than the control group, and recall describes how much of the test group $P$ can cover.
For example, the pattern $(e_2)$ has a precision of 0.75 because it occurs in 4 traces in total, 3 of which are in the test group, i.e., it is 75\% specific to the test group.
It also has a recall of 0.6 because it occurs in 3 out of 5 traces in the test group, i.e., it covers 60\% of the test group.
The $\fone$ score of this pattern is simply the harmonic mean of the two quantities, 0.67.
\name{} computes the $\fone$ scores of all patterns and returns the list of 
all patterns ranked by $\fone$ scores. 

In our example (\figref{example}), the pattern $(e_2, e_3)$ is the highest ranked.
This makes sense intuitively, as in this contrived example, the pattern occurs disproportionately and consistently among users who experienced the bug compared to others -- 3 out 5 users who experienced the bug versus none among the 5 other users.
A software engineer debugging the reports can infer that events $e_2$ and $e_3$, occurring in that order, are likely the source of a bug and deserve close inspection.
In Section~\ref{sec:evaluation}, we evaluate the accuracy and utility of such patterns in practice. 

An astute reader might have noted that the pattern $(e_5, e_7)$ offers an interesting alternative explanation of the bug, but is ranked lower than patterns involving $e_2$ and $e_3$ that might seem redundant with respect to the top ranked pattern.
We will come back to this in Section~\ref{subsec:redundancy}.

%% file: rca-framework.tex

\section{RCA Framework}
\label{sec:technical}

In this section, we present technical details about scalable statistical RCA using \name{}.

\subsection{Preliminaries}
\label{subsec:language}

The basic unit of our RCA method is an {\em event}, 
which records some telemetry about the state of the execution of an app at some point in time.
For instance, an event could capture the fact that ``the user uploaded a photo'', or ``the OS raised a low memory alarm''.
Developers typically track events by instrumenting the app code with logging statements.
Our model is agnostic to the actual content of an event, such as the text of the button tapped\footnote{Section~\ref{sec:practical} discusses a special case of handling events with numeric data.}.
We will denote events by $e_1, e_2, \ldots$, 
and assume that they come from a 
finite vocabulary $E$.
An execution of an app naturally creates a chronological sequence of events,
or a \emph{trace}:

\begin{definition}[Trace]
\label{def:trace}
A trace is a contiguous sequence of events $\tau = (e_1, \ldots, e_k)$ where event $e_i$ is followed by event $e_{i+1}$ during the execution of the app.
\end{definition}
Informally, a trace can be thought of as the sequence of instrumented program points of interest that were visited during the execution of the app.

The goal of our method is to extract \emph{patterns} in traces that 
are indicative of the root cause of the bug.
To do this, we need to first define a language of patterns.
Our choice for such a language is primarily driven by how we intend to deploy \name{}.
In our setting, the patterns are intended to be presented to developers 
who can make further judgements, often in a time-sensitive manner to mitigate the bug quickly.
It is, thus, important that our patterns be amenable for human interpretability and scalability.
Next, it is also desirable that patterns preserve temporal aspects of traces arising from app executions.
Our choice of {\em sequential patterns}~\cite{agrawal95mining}, in fact, meets all these desired properties.

\begin{definition}[Pattern]
\label{def:pattern}
A sequential pattern, hereafter simply ``pattern'', is a possibly non-contiguous sequence of events $p = (e_1, \ldots, e_k)$, where event $e_i$ is {\em eventually} followed by $e_{i+1}$.
\end{definition}

\ignore{
Defining the pattern language to be close \edward{?} to traces makes it highly interpretable for developers who are already familiar with traces.
It also enables us to tap into algorithms from the data mining community for efficiently enumerating the language, as we will see next.
}

\subsection{Scalable Enumeration of Patterns}
\label{subsec:scalable}

To realize our objective of extracting meaningful patterns from traces, one naive way is to enumerate all possible patterns over our vocabulary (say, up to a fixed length), and count how many traces match against each pattern.
This is clearly not a feasible approach due to combinatorial explosion.
We instead leverage ideas from the pattern mining literature which overcomes this problem
and discovers meaningful patterns while avoiding the combinatorial blowup. 
In this section, we will formalize these notions.

We say that a sequence $\alpha = (a_1, a_2, \ldots, a_n)$ is a \emph{subsequence} of another sequence  $\beta = (b_1, b_2, \ldots, b_m)$, denoted $\alpha \sqsubseteq \beta$, if there exist indices $1 \leq j_1 < j_2 < \cdots < j_n \leq m$ such that $a_1 = b_{j_1}, a_2 = b_{j_2}, \ldots, a_n = b_{j_n}$.
A trace $t$ is said to contain pattern $p$ (or, pattern $p$ appears in trace $t$), if $p \sqsubseteq t$.

\begin{definition}[Support]
\label{def:support}
Given a set of traces $S = \{t_1, t_2, \ldots, t_n\}$, the support of a sequence $\alpha$ in $S$ is the number of traces in $S$ that contain $\alpha$:
$$
support_S(\alpha) = |~ \{t_i~|~t_i \in S \wedge \alpha \sqsubseteq t_i\} ~|
$$
\end{definition}
This allows us to specify a minimum support threshold $\minsup$, a positive integer, such that a sequence $\alpha$ is considered a pattern in $S$ only if $support_S(\alpha) \geq \minsup$.

The data mining community has extensively studied the problem of mining patterns from trace-like data, 
and several principled approaches based on the {\em a priori} property~\cite{agrawal95mining,srikant1996mining,Pei2004,wang2004bide} have been developed.
The {\em a priori} property states that the support of a sequence in $S$ is bounded by the support of any of its subsequences in $S$.
This property is exploited in the {\em PrefixSpan} algorithm~\cite{Pei2004}, which we use in \name{}.
We next describe the intuitions behind this algorithm and how we adapt it to scale to our setting.


\begin{definition}[Suffix with respected to pattern]
\label{def:prefix}
Let $t$ be a sequence and let $\alpha$ be a pattern.
We say that $\gamma$ is a suffix of $t$ with respect to $\alpha$ 
if 
there is a prefix $t'$ of $t$ such that $\alpha \sqsubseteq t'$ and $t = t' \cdot \gamma$.
We say $\gamma$ is the maximal suffix of $t$ with respect to $\alpha$ if it is the longest
such suffix.
\end{definition}

\begin{figure}
\begin{center}
\begin{small}
\begin{tikzpicture}[
emptynode/.style={rectangle},
squarednode/.style={rectangle, draw=cyan!60, fill=cyan!5, very thick, minimum size=5mm, align=center},
]
\tikzset{cross/.pic={\draw[rotate=45, line width=0.5mm] (-#1,0) -- (#1,0);\draw[rotate=45, line width=0.5mm] (0,-#1) -- (0, #1);}}

\node[squarednode](alpha){$\alpha = (a_2, a_3)$ \\ $S|_\alpha = \{\gamma_1, \gamma_2, \gamma_3\}$ \\ $support(\alpha) = 3$};
\node[squarednode](beta1)[below left=0.2cm and -1cm of alpha]{$\beta_1 = (a_2, a_3, a_4)$ \\ $S|_{\beta_1} = \{\gamma_1', \gamma_2'\}$ \\ $support(\beta_1) = 2$};
\node[squarednode](beta2)[below right=0.2cm and -1cm of alpha]{$\beta_2 = (a_2, a_3, a_5)$ \\ $S|_{\beta_2} = \{\}$ \\ $support(\beta_2) = 0$};
\node[emptynode](next1)[below left=-0.3cm and 0.5cm of beta1]{$\ldots$};
\node[emptynode](next2)[below right=-0.4cm and 0.5cm of beta2]{};

\draw[->] (alpha.west) -| (beta1.north);
\draw[->] (alpha.east) -| (beta2.north);
\draw[->] (beta1.west) -| (next1.north);
\draw[->] (beta2.east) -| (next2.north);
\path(next2.north) pic[red] {cross=4pt};

\end{tikzpicture}
\end{small}
\end{center}
\caption{Tree-traversal of PrefixSpan when exploring pattern space.}
\figlabel{prefixspan}
\end{figure}
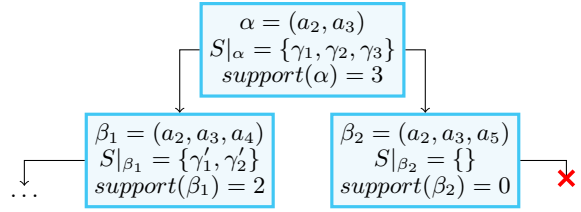

PrefixSpan regards pattern mining as a tree-traversal problem, where the nodes are patterns and the parent of a node is its prefix, as shown in \figref{prefixspan}.
It works by enumerating the prefix of patterns, starting from the empty pattern.
At each node (pattern) in the tree, it expands the children by appending one event to the pattern.
At each node, it also maintains the set of maximal suffixes from traces in $S$ that contain the pattern as the prefix, called the ``projected database''.


\begin{definition}[Projected Database]
\label{def:projected}
Let $\alpha$ be a pattern in $S$.
The $\alpha$-projected database, denoted $S|_\alpha$, is the set of maximal suffixes of sequences in $S$ with respect to the pattern $\alpha$. 
\end{definition}

Now, the crux of the algorithm is that the support of any node's (pattern's) children can be obtained from the projected database associated with the node, rather than the original set of traces $S$.
The {\em a priori} property guarantees soundness -- that no trace not in the projected database would contain any of the pattern's children.
Since suffixes get shorter as the pattern gets longer, the projected databases keep shrinking with the depth of the tree, making the algorithm highly efficient.
Moreover, if the projected database at any node becomes empty, the entire subtree of patterns under the node can be pruned away, as shown in \figref{prefixspan}.

\begin{lemma}[Projected Database]
\label{lemma:projected}
Let $\alpha$, $\beta$ and $\delta$ be patterns such that $\beta = \alpha \cdot \delta$.
We have,
\begin{itemize}
    \item[1.] $S|_\beta = (S|_\alpha)|_\delta$
    \item[2.] $support_S(\beta) = support_{S|_\alpha}(\delta)$
    \item[3.] the size of $S|_\alpha$ cannot exceed that of $S$
\end{itemize}
\end{lemma}

\input{extract_patterns}

Taking advantage of Lemma~\ref{lemma:projected}, \name{} runs a recursive divide-and-conquer PrefixSpan algorithm to mine patterns in a set of traces $S$.
Algorithm~\ref{algo:prefixspan} shows the pseudocode of this method.
It is invoked with \textsc{ExtractPatterns}$(\emptyseq, S, \minsup)$ where $\emptyseq$ is the empty sequence.
We assume that the procedure \textsc{ProjectedDatabase}($S$, $\alpha$) computes the $\alpha$-projected database of $S$ as defined in Definition~\ref{def:projected}.
This is a computationally expensive operation, but interested readers can refer to~\cite{Pei2004} for practical implementation tricks to speed it up.
The algorithm finally returns a set of tuples of the form $(p, \{i_1, \ldots i_n\})$ where $p$ is a pattern that appears in at least $\minsup$ traces in $S$, and $ \{i_1, \ldots i_n\}$ are the IDs of the traces in $S$ that contain $p$.
The cardinality of this set is $support_S(p)$, but we will see later in Section~\ref{sec:practical} why returning the trace IDs themselves is useful.

\subsection{Statistical Model}
\label{subsec:statistical}

\begin{algorithm}[tb]
\caption{\name{}'s algorithm for RCA}
\label{algo:statistical}
\begin{algorithmic}[1]
\begin{small}
\Procedure{MinesweeperRCA}{$T$, $C$, $\minsup$}
\State \Comment{$T$ and $C$ are the test and control groups, respectively}
\State result $R \gets \emptyset$
\State $patterns_T \gets \textsc{ExtractPatterns}(\emptyseq, T, \minsup)$ 
\State $patterns_C \gets \textsc{ExtractPatterns}(\emptyseq, C, \minsup)$ 
\State \vspace{-5mm}\begin{equation*}
\hspace{4mm}patterns \gets \bigcup
\begin{cases}
\{(p, ts, cs)\} & (p, ts) \in patterns_T \wedge \\ & \hspace{3mm} (p, cs) \in patterns_C\} \\
\{(p, ts, \emptyset)\} & (p, ts) \in patterns_T \wedge \\ & \hspace{3mm} (p, \cdot) \notin patterns_C\}
\end{cases}
\end{equation*}
\For{(pattern $p$, test sup. $ts$, control sup. $cs$) {\bf in} $patterns$}
\State $precision \gets |ts| ~/~ (|ts| + |cs|)$
\State $recall \gets |ts| ~/~ |T|$
\State $\fone\_score \gets \textsc{HarmonicMean}(precision, recall)$
\State $R \gets R \cup \{(p, precision, recall, \fone\_score)\}$
\EndFor
\State $R_{ranked} \gets \textbf{sorted}(R, \lambda (\cdot, \fone\_score): -\fone\_score)$
\State \Return $R_{ranked}$
\EndProcedure
\end{small}
\end{algorithmic}
\end{algorithm}

We have so far presented a general algorithm to extract frequent patterns in a group of traces.
Suppose that we are given a test group $T = \{t_1, t_2, \ldots t_n\}$ where the bug was encountered.
We can immediately invoke \textsc{ExtractPatterns} on $T$ with some minimum support threshold to get the most common patterns in $T$.
However, in order
to isolate patterns that are {\em distinctive} to $T$, \name{} also accepts a set of traces $C$, the control group, where the bug was not encountered.
%
Algorithm~\ref{algo:statistical} provides the pseudocode of \name{}'s RCA method.
It first extracts patterns in both $T$ and $C$, which returns the supporting traces for each pattern in the two groups.
Then, for each pattern $p$, it computes the following two quantities.

\begin{definition}[Precision]
\label{def:precision}
The precision of a pattern $p$ is defined as the probability of a trace $t$ being in the test group $T$, given that $p$ appears in $t$.
\begin{align*}
Precision(p) &= \text{Pr} (t \in T ~|~ p \sqsubseteq t)\\
&= \frac{support_T(p)}{support_T(p) + support_C(p)}
\end{align*}
\end{definition}

\begin{definition}[Recall]
\label{def:precision}
The recall of a pattern $p$ is the probability that $p$ appears in a trace $t$, given that $t$ is in the test group $T$.
$$
Recall(p) = \text{Pr} (p \sqsubseteq t ~|~ t \in T) = \frac{support_T(p)}{|T|}
$$
\end{definition}

Together, the precision and recall of a pattern $p$ quantify how distinctive $p$ is to the test group as opposed to the control group.
To work with a single measure, \name{} computes their {\em $\fone$-score}, which is simply the harmonic mean of the two.
Essentially, the higher the $\fone$-score of a pattern, the more powerful it is in isolating the test group from the control group.
\name{} computes the $\fone$-score of each pattern 
and finally returns a ranked list of patterns sorted, in descending, by their $\fone$-scores.

%% file: extract_patterns.tex

\begin{algorithm}[tb]
\begin{small}
\caption{Algorithm for extracting patterns from traces}
\label{algo:prefixspan}
\begin{algorithmic}[1]
\Procedure{ExtractPatterns}{$\alpha$, $S|_\alpha$, $\minsup$}
\State \Comment{Invoked as \textsc{ExtractPatterns$(\emptyseq, S, \minsup)$}}
\State result $R \gets \emptyset$
\For{event $e$ {\bf in} $E$}
\State $\beta \gets \alpha \cdot e$
\If{$support_{S|_\alpha}(\beta) < \minsup$}
\State \textbf{continue}
\EndIf
\State $R \gets R \cup \{(\beta, \{i~|~t_i \in S|_\alpha \wedge \beta \sqsubseteq \alpha \cdot t_i\})\}$
\State $S|_\beta \gets \textsc{ProjectedDatabase}(S|_\alpha, \beta)$
\State $R \gets R \cup \textsc{ExtractPatterns}(\beta, S|_\beta, \minsup)$
\EndFor
\State \Return $R$
\EndProcedure
\end{algorithmic}
\end{small}
\end{algorithm}

%% file: rca-practice.tex

\section{Practical Considerations for RCA}
\label{sec:practical}

In this section, we discuss technical challenges that arise for performing RCA in a practical setting.

\subsection{Mitigating Redundant Patterns}
\label{subsec:redundancy}
The algorithm in Section~\ref{sec:technical} is efficient at extracting and ranking patterns that point towards the root cause.
However, it can sometimes return {\em redundant} patterns -- patterns that are similar in explaining the bug.
For instance, in the example in \figref{example}, the patterns $(e_2)$ and $(e_3)$ point to a similar root cause as the top pattern $(e_2, e_3)$.
In contrast, the pattern $(e_5, e_7)$ offers an alternative explanation of the bug, but is lower ranked than these patterns.
In practice, engineers would want to see patterns that explain varying facets of the bug rather than redundant patterns that differ only slightly.


To quantify if two patterns are redundant, \name{} computes a similarity score between them based on their supporting traces.
Suppose $\alpha$ and $\beta$ are two patterns returned by Algorithm~\ref{algo:statistical}, and $ts_\alpha$ and $ts_\beta$ are the IDs of the traces in $T$ that support $\alpha$ and $\beta$, respectively. 
\name{} uses {\em Jaccard similarity} between $ts_\alpha$ and $ts_\beta$, defined as:
$
similarity(\alpha, \beta) = |ts_\alpha \cap ts_\beta| / |ts_\alpha \cup ts_\beta|
$.
This is a value between 0 and 1 such that the higher it is the more in common are traces in which both patterns appear, indicating that one of the patterns is redundant.

\name{} lets engineers pick a similarity threshold they are comfortable with, such that two patterns are considered redundant if their similarity score is above the threshold.
Then, for each pattern, it computes the group of patterns that are similar to it beyond the threshold -- this ends up forming {\em clusters} of patterns that are being supported by similar sets of traces.
From each cluster, it picks the pattern with the highest $\fone$-score.
If there is a tie, it picks the longer pattern, but this is simply a heuristic choice.

\figref{redundancy}(a) shows the patterns from \figref{example} along with their similarity scores -- only with respect to $(e_2, e_3)$ and $(e_5, e_7)$ for brevity.
With a similarity threshold of 0.6, there would be two clusters of patterns, from which $(e_2, e_3)$ and $(e_5, e_7)$ will be selected, respectively.
As illustrated in~\figref{redundancy}(b), such a pre-processing step eliminates many of the redundant patterns and the result is more succinct in explaining varied aspects of the bug.
In practice, engineers set a similarity threshold of 0.9, which results in 85-90\% of (redundant) patterns dropped.


\begin{figure}
{\small
\begin{tabular}{c}
\begin{tabular}{|c|c|c|c|c|}
\hline
{\bf Pattern} & {\bf $\fone$} & {\bf Supporting} & \multicolumn{2}{c|}{\bf Similarity w.r.t.} \\
\cline{4-5}
& {\bf score} & {\bf traces} & $(e_2, e_3)$ & $(e_5, e_7)$ \\
\hline
    $(e_2, e_3)$ & 0.75 & $\{t_1, t_2, t_3\}$ & 1.0 & 0 \\
    $(e_2)$ & 0.67 & $\{t_1, t_2, t_3\}$ & 1.0 & 0 \\
    $(e_3)$ & 0.6 & $\{t_1, t_2, t_3\}$ & 1.0 & 0 \\
    $(e_1, e_2, e_3)$ & 0.57 & $\{t_1, t_2\}$ & 0.66 & 0 \\
    $(e_5, e_7)$ & 0.57 & $\{t_4, t_5\}$ & 0 & 1.0 \\
\hline
\end{tabular}
\\
(a)
\\
\begin{tabular}{|c|c|c|c|c|c|}
\hline
{\bf Pattern} & \multicolumn{2}{c|}{\bf Support} & {\bf Prec-} & {\bf Recall} & {\bf $\fone$} \\ \cline{2-3}
& {\bf Test} & {\bf Control} & {\bf ision} && {\bf Score} \\
\hline
    $(e_2, e_3)$ & 3 & 0 & 1.0 & 0.6 & 0.75 \\
    $(e_5, e_7)$ & 2 & 0 & 1.0 & 0.4 & 0.57 \\
\hline
\end{tabular}
\\
(b)
\end{tabular}
}
\caption{(a) Similarity scores of patterns from \figref{example}, (b) Patterns from \figref{example} with a similarity threshold of 0.6.}
\figlabel{redundancy}
\end{figure}

\subsection{Handling Numeric Data in Traces}
\label{subsec:numeric}

Many valuable signals come from continuous sources, for example the amount of memory the Facebook app consumed before a crash, or the number of the times the app was opened in the last 24 hours.
However, numeric data drawn from continuous distributions presents a problem for \name{} -- it violates our assumption that events come from a finite discrete vocabulary $E$.
Naively treating each occurrence of a continuous value as a unique event would likely explode the size of $E$, leading to scalability issues.

\ignore{
Firstly, unless the data is highly structured, it is unlikely that any values of a continuous feature will re-appear frequently enough to meet even low minimum support thresholds and thus be mined. Secondly, even if certain values of a continuous feature do have enough support, the resulting patterns formed with these features are hard for an engineer to gain information from. For example, if Minesweeper detects a problematic pattern in the test group with an app memory usage of 10 megabytes, it is hard to determine whether this is abnormally high or low memory usage without domain knowledge.

An easy solution to the first problem, supplying an increased amount of data and lowering the minimum support threshold, significantly increases the runtime of the algorithm and still produces hard-to-decipher patterns.}

To deal with such problems involving continuous data, we use {\em discretization}, a technique common to many algorithms across the data-mining and machine learning fields.
We add a preprocessing step to \name{} that detects and discretizes continuous data.
More specifically, if some feature has a continuous distribution with minimum value $a$ and maximum value $b$, we produce an increasing sequence of endpoints $a < x_1 < x_2 \ldots < x_n < b$, and partition the domain $[a, b]$ into ``bins'' $[a,x_1], (x_1,x_2], (x_2,x_3], \ldots, (x_n,b]$, where $n$ is user specified or calculated in accordance with the size of the data.
Then, given a specific value $v$ of the continuous feature, we can simply replace it with the bin $(x_i, x_{i+1}]$ such that $x_i < v \leq x_{i+1}$.
We can then feed the binned (discretized) inputs to \name{} as the bins are finite and discrete.

This procedure requires some design decisions, such as choosing the number of bins $n$, and placing the endpoints $x_i$ such that good patterns can be mined.
We explored and tested several combinations for these choices on validation data sets.
For selecting the number of bins, standard heuristics such as Sturges' rule~\cite{sturges} and Freedman-Diaconis' rule~\cite{Freedman81onthe} can be applied, which compute $n$ based on either the size of the data set or a simple statistic like the inter-quartile range (IQR).

For deciding how the endpoints $x_i$ should be placed, we explored a few strategies:
\begin{enumerate*}[label=(\roman*)]
\item equal proportion, which ensures each bin gets the same number of points in the data set,
\item equal width, which ensures the range of each bin, i.e., $x_{i+1}-x_i$, is the same, and
\item k-bins, a clustering algorithm that makes values in each bin have the same nearest center of a 1-dimensional cluster.
\end{enumerate*}
Overall, most combinations of the two choices we made produced similar and reasonable splits, which led to meaningful patterns mined from \name{}, with the exception of the equal-width strategy as it does not consider the distribution of the data.
We allow users to pick any strategy when invoking \name{}, as the problem domain may make one of these choices more useful than others.

\ignore{
All of the above techniques draw from unsupervised approaches to discretization, which may in some sense be losing more information compared to supervised discretization techniques. However, Minesweeper makes use of the test and control group split later on during statistical isolation. For this reason, the loss of information during unsupervised discretization here is less worrisome. In practice, we found that the available options for users produced satisfactory results.
}



%% file: rca-facebook.tex

\section{Root Cause Analysis at Facebook}
\label{sec:rca-fb}

\begin{figure*}
    \centering
    \includegraphics[width=2.1\columnwidth]{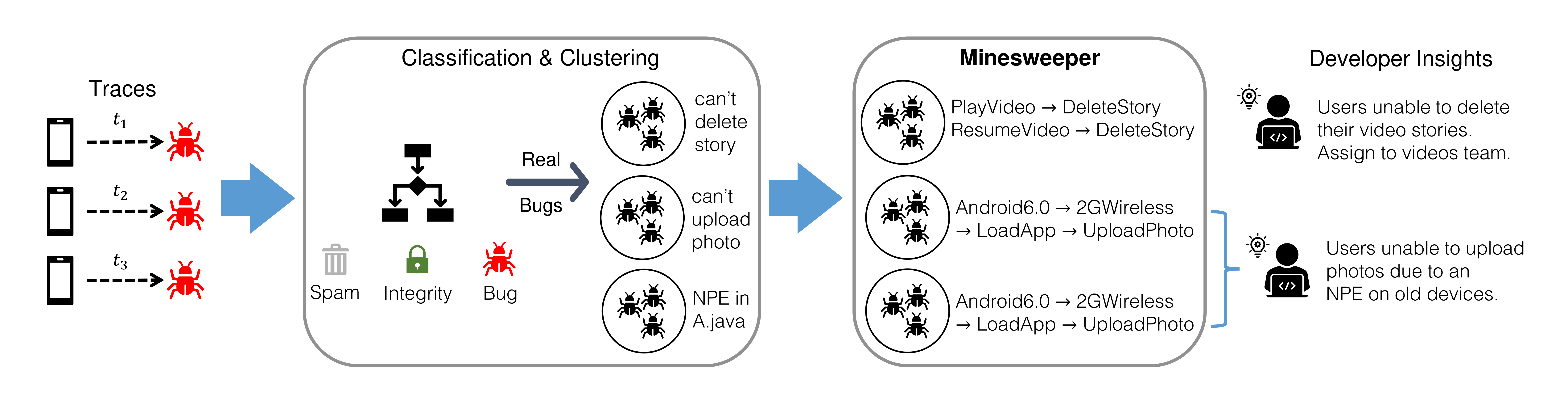}
    \vspace{-0.3in}
    \caption{Root Cause Analysis of bugs and crash reports at Facebook using Minesweeper}
    \figlabel{rca-fb}
    \vspace{-0.2in}
\end{figure*}
\ignore{
\begin{figure}
    \centering
    \includegraphics[width=\columnwidth]{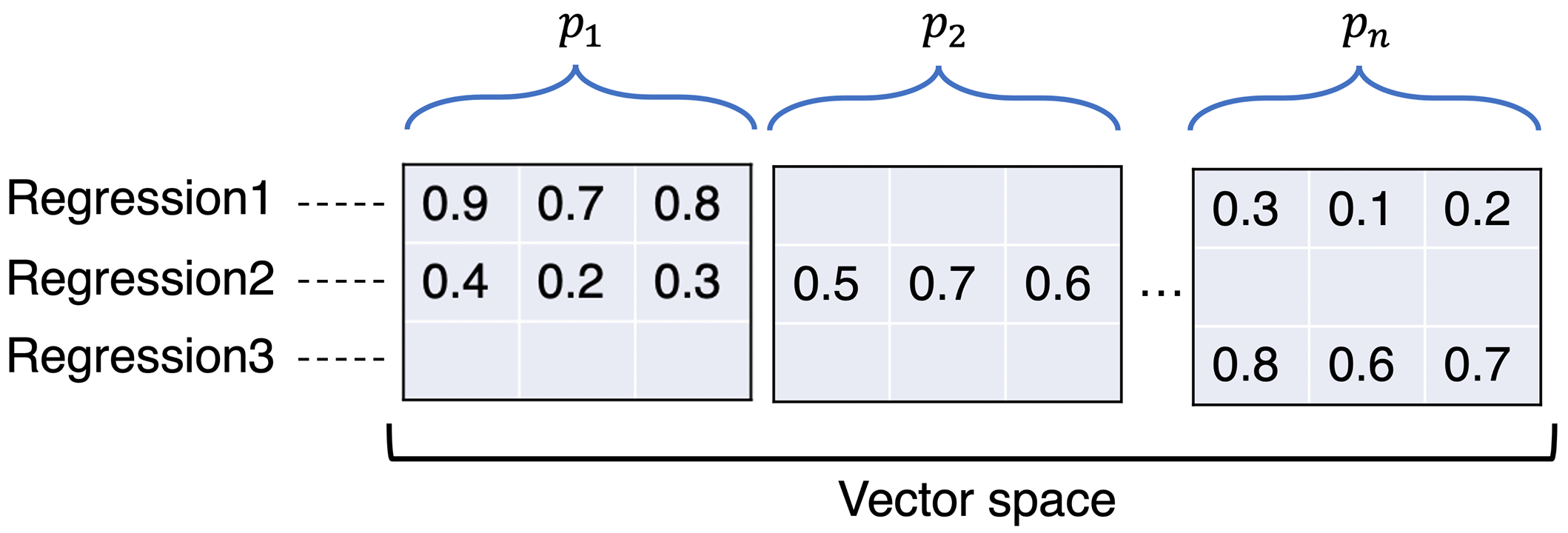}
    \caption{Encoding regressions into a vector space using \name{} patterns. Each pattern's triplet of precision, recall and f1-score is flattened into a vector for the regression.}
    \figlabel{embedding}
\end{figure}
}

In this section, we examine how \name{} patterns can be used in an industrial setting, and discuss how RCA of bugs and crashes at Facebook can be enhanced by \name{}.

\subsection{Representing Regressions using Patterns}
\label{subsec:regressions}

A practical application of \name{} in an industrial setting like Facebook is to help engineers root cause and diagnose {\em regressions} -- sudden spikes in a group of crash or bug reports.
With the complexity of Facebook's apps and the product release cycle, it is common for multiple regressions to take place simultaneously, especially after a new version is released.
In this real time setting, engineers have to be provided with high quality insights about each regression that would help them debug the issue.
\name{} patterns serve this purpose well, as they provide a kind of ``signature'' for the regression, capturing its most distinctive properties.

To encode regressions, \name{} first runs Algorithm~\ref{algo:statistical} on all regressions, with the control group being non-regressing traces, and computes the set $P = \{p_1, p_2, \ldots, p_n\}$ of all patterns present in all regressions.
A given regression is then represented using a vector $v = (u_1, u_2, \ldots, u_d)$ in some high-dimensional vector space $d$.
To compute the co-ordinates of $v$ for a regression, \name{} uses the precision, recall, and $\fone$-score of patterns in $P$.
Specifically, pattern $p_i$'s precision, recall and $\fone$-score go into co-ordinates $(u_{j-2}, u_{j-1}, u_j)$, respectively, where $j=3i$.
%
%
If $p_i$ is not relevant to the given regression, its corresponding triplet of co-ordinates is set to 0.

This encoding allows computing an important insight about regressions -- how related they are -- using standard vector space distances such as {\em cosine distance}.
Specifically, if the distance between two regressions is small, there is likely to be some relationship between them based on their patterns (root cause).
This enables linking together regressions, even cutting across traditional debugging boundaries.
For instance, not all users report bugs that they encounter, but \name{} can potentially link the bug reports to a crash -- which is automatically reported -- allowing engineers to assess the true impact of user-facing bugs.
As another example, a bug can manifest as two different crashes, say, out-of-memory or killed by the OS, depending on features of the device.
These would appear as two different types of regressions, but using their vector encodings \name{} can link them together and help developers debug them.
In Section~\ref{sec:evaluation} we conduct an experiment with this type of encoding to measure the accuracy of \name{} in identifying the root cause of regressions.


\subsection{\name{} in Facebook's RCA workflow}
\label{subsec:fb-workflow}

Facebook receives many bug reports and crashes on a daily basis, ranging from users not being able to delete their story, to the app crashing due to a null pointer exception (NPE).
When a crash or bug is encountered, data is sent to Facebook via pre-configured logging built into the app or through user-submitted reports.
The information sent includes details such as the time and type of crash, metadata such as the version of the Facebook app being used, a stack trace if available, and traces of events that led up to the bug or crash, as shown in \figref{rca-fb}. 
In the case of bug reports, user-provided information can be useful but often is noisy and hard to decipher.

After many reports are logged, various classifiers attempt to classify and cluster the reports into groups.
These aggregated groups each receive their own label that corresponds to the common symptom that all the reports in the group face, for example, users not being able to delete their story.
Various alerts and metrics are attached to each group to notify engineers about anomalous behaviors in the group, such as if volume of incoming reports suddenly starts spiking.

Generally, if any such alerts fire, the first line of defense will be an ``on-call engineer'', who has the task of investigating the problem and triaging it to appropriate engineers for fixing it.
This is a time-sensitive task especially due to Facebook's scale of operations, and a non-trivial one, as useful debugging signals -- such as a stack trace -- are often absent in bug reports.
In these cases, \name{} can help take the groups of bug reports and extract useful patterns that point to the root cause.

Specifically, on-call engineers can take hundreds or thousands of reports from the problematic group to be the test group.
For the control group, they may opt to select reports from another group of errors, or randomly sample non-buggy or non-crashing sessions. 
They then feed both groups to \name{}, and minutes later will have a ranked list of patterns that are distinctive to the test group of errors.
These patterns oftentimes provide a strong signal in pointing engineers to the correct files and modules to look for the problematic behavior.
From there, it is easy for the issue to be triaged to the team owning the file or module from which the error occurs, resulting in a fast fix.
Sometimes, \name{} can also relate seemingly disparate errors that come in through different error-reporting systems, using their root cause. 

%% file: evaluation.tex

\ignore{
\begin{figure*}[t]
    \centering
    \begin{tabular}{ccc}
    \includegraphics[width=0.6\columnwidth]{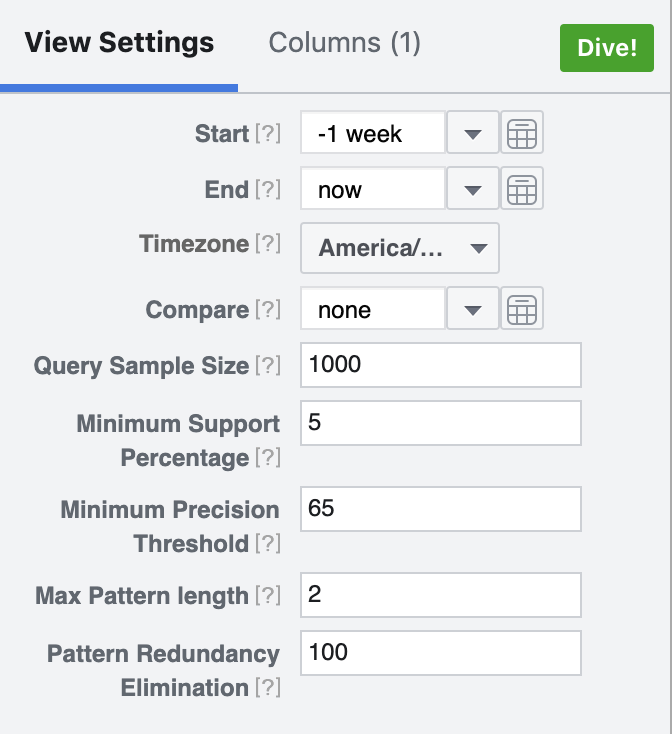}
    & \hspace{0.2in}
    & \includegraphics[width=0.8\columnwidth]{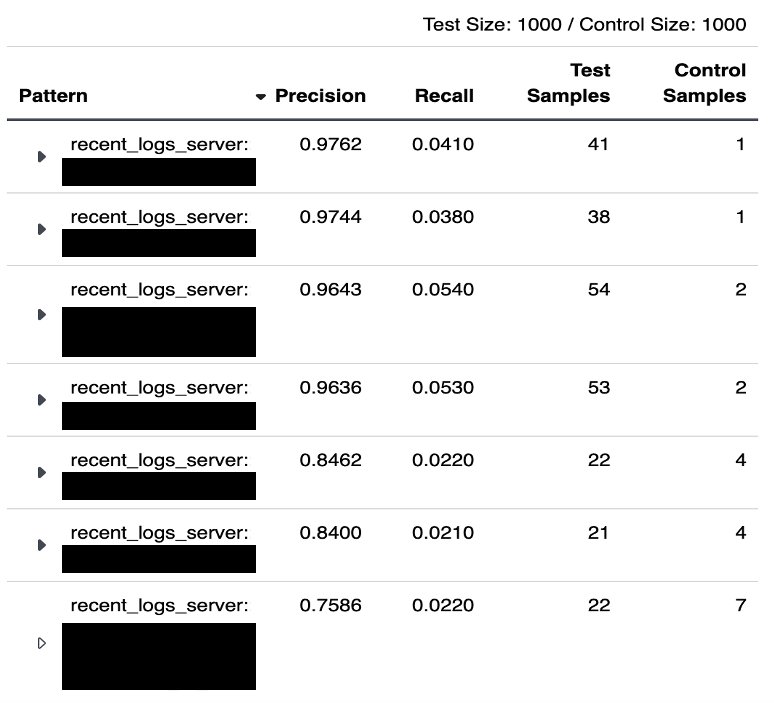}\\ 
    (a) & & (b)
    \end{tabular}
    \caption{\name{} UI with patterns.}
    \figlabel{ui}
\end{figure*}
}


\begin{figure}
    \centering
    \includegraphics[width=0.95\columnwidth]{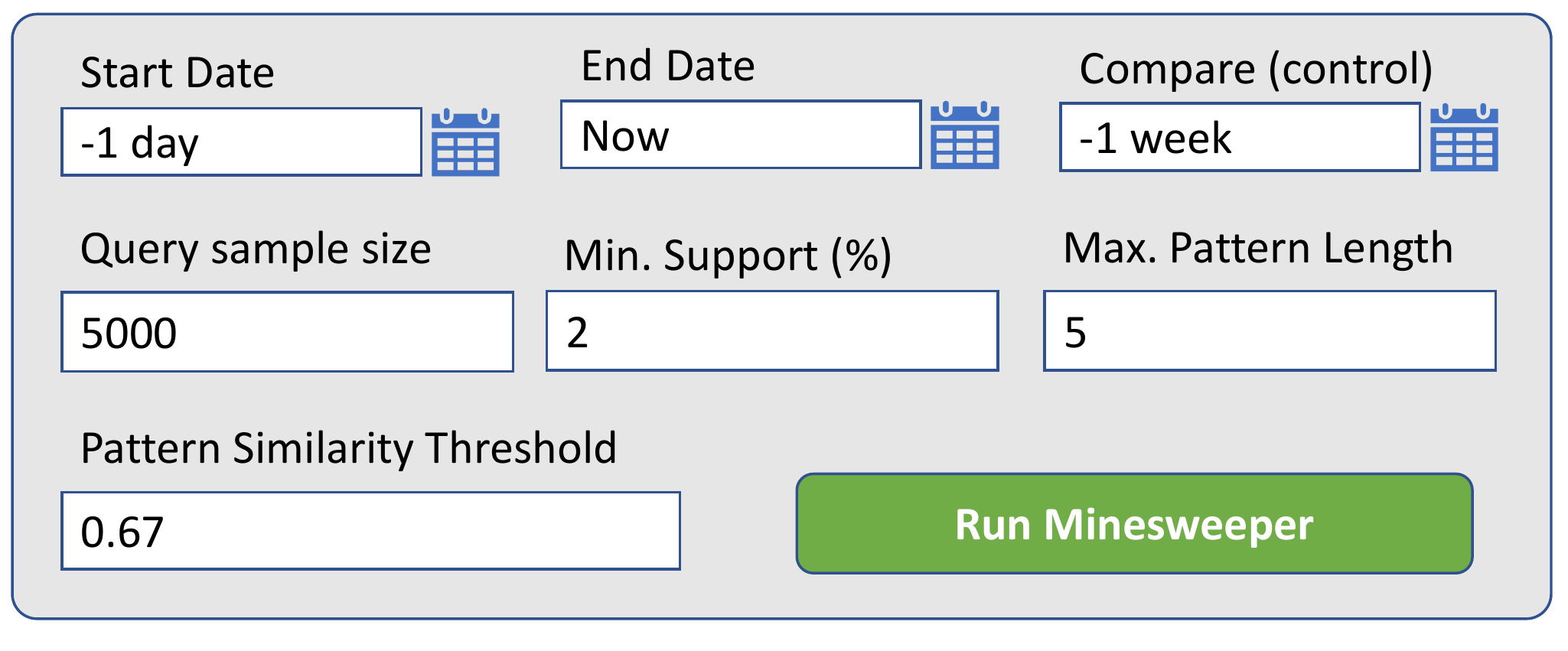}
    \vspace{-0.2in}
    \caption{Mockup of \name{}'s UI.}
    \figlabel{ui}
\end{figure}

\begin{figure*}
\subfigure[Number of traces]
{
    \includegraphics[width=5.7cm, height=4cm]{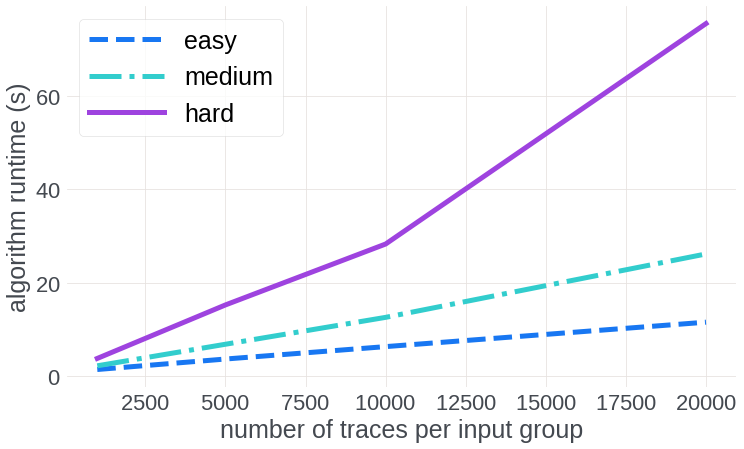}
    \figlabel{scalability_traces}
}
\subfigure[Average trace length]
{
    \includegraphics[width=5.7cm, height=4cm]{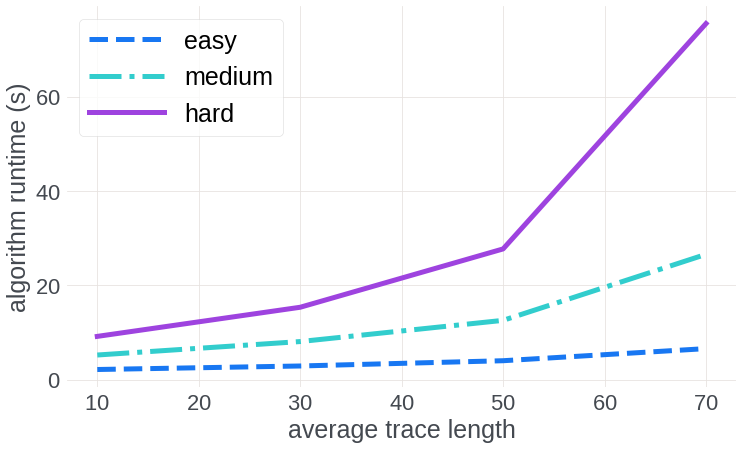}
    \figlabel{scalability_avg_len}
}
\subfigure[Minimum support threshold]
{
    \includegraphics[width=5.7cm, height=4cm]{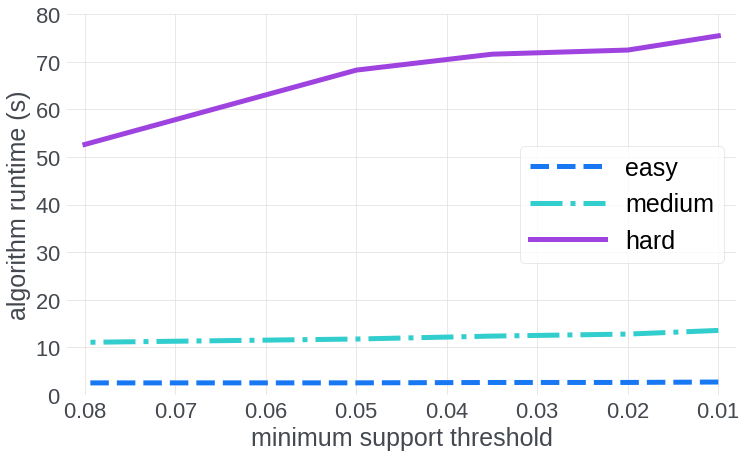}
    \figlabel{scalability_min_sup}
}
\vspace{-0.15in}
\caption{\name{}'s scalability along various dimensions: (a) number of traces \revision{per input group}, (b) average length of traces, (c) minimum support threshold.}
\figlabel{eval-scalability}
\vspace{-0.2in}
\end{figure*}


\section{Evaluation}
\label{sec:evaluation}

We implemented \name{} as described in previous sections, and have evaluated it on developer operations at Facebook. In this section, we present these results.

\subsection{Implementation}
\label{subsec:implementation}


\name{} is primarily written in Python, using the Scikit library for numeric operations, and the data-mining library SPMF~\cite{spmf} for extracting patterns from traces.
\figref{ui} shows a mockup of a user interface for \name{}.
Given a time period, this interface can be used to query test and control group traces from Facebook's bug report data store.
Engineers can also specify various parameters for the algorithm such as $\minsup$ before invoking \name{}.
The core system accepts the traces and parameters, discretizes continuous data as described in Section~\ref{subsec:numeric}, runs Algorithm~\ref{algo:statistical}, and finally removes redundant patterns as described in Section~\ref{subsec:redundancy}.
The final ranked list of patterns is returned and displayed as shown in the table in \revision{\figref{redundancy}}.

\subsection{Scalability along various dimensions}
\label{subsec:evel-scalability}


We evaluate the runtime of \name{} along several important dimensions to show that the algorithm is useful for practical purposes.
We test a grid of input parameters, with the input data consisting of telemetry from the Facebook Android app.
The traces in this data have been filtered and/or pre-processed by other tools not related to \name{}.
In this experiment, we test how runtime scales with the number of traces in each input group, the length of each trace, and the minimum support threshold.
We separate the groups of trials into three groups of difficulty for \name{} to process, namely `easy', `medium', and `hard', corresponding to how long we expect \name{} to take to run each of the jobs. 

In the `hard' runs, the test and control groups each contain 20,000 traces, the minimum support threshold is 0.01, and the median trace length is 70.
`Medium' runs have 10,000 traces per group, a median support threshold of 0.0275, and a median trace length of 40.
`Easy' runs have a median of 3000 traces per group, a minimum support threshold of 0.05, and a median trace length of 20 events.
In all runs, the maximum pattern lengths is restricted to 5, to support human interpretability.

\figref{eval-scalability}(a) shows that \name{}'s runtime scales linearly with the number of traces.
Although each additional trace can potentially introduce new combinatorial behavior among events, we observe that it does not happen in practice.
Typically, a relatively fewer number of traces is sufficient to `cover' most patterns appearing in the group, as the traces share a lot of homogeneity.
\figref{eval-scalability}(b) shows that the effect of increasing trace length on runtime is more exponential in nature.
This arises from the fact that increasing trace length pulls in more events in each trace, combinatorially increasing the number of patterns.
This also shows that there is not much homogeneity to leverage on among events within a single trace.
Finally, \figref{eval-scalability}(c) shows that lowering the minimum support threshold has a close to linear effect on the runtime.
Though this contrasts with previous studies on PrefixSpan and related a-priori based mining algorithms~\cite{wang2004bide}, which show exponential increases in runtime due to minimum support settings.
We observe this quasi-linear relationship in practice for the same reason number of traces scales linearly with runtime.

\ignore{
\figref{eval-scalability}(a) shows, somewhat surprisingly, that the runtime of Minesweeper scales fairly linearly with the number of traces supplied. We hypothesize this is because the event vocabulary of our test data is almost entirely `covered' already by relatively few traces, i.e. a great majority of the patterns will be picked up whether we supply 1000 or 50000 traces. Because the event vocabulary is already 'covered', the marginal trace added to the input of the algorithm does not increase the amount of work that the algorithm has to do on previously existing inputs. 

\figref{eval-scalability}(b) shows that as the average length of a trace increases, the runtime of the algorithm increases exponentially. This aligns with intuition that an increase in the trace length increases the number of possible patterns exponentially, which means in the worst case that pattern mining, statistical isolation, and pattern redundancy steps have to each do exponentially more work. As opposed to adding a number of traces, adding new columns necessarily expands the vocabulary size, and thus does not enjoy linear scaling.

As shown in \figref{eval-scalability}(c), as the minimum support threshold is lowered, the runtime of Minesweeper appears to increase exponentially. This is in line with previous studies on PrefixSpan and related a-priori based mining algorithms \edward{citations}. Because we build Minesweeper on top of these mining algorithms, we expect that the performance cannot asymptotically scale better exponentially.
}

\subsection{Use cases and qualitative feedback}
We outline two cases where \name{} aided the debugging process of an error at Facebook.
The first comes from an error from JSON decoding that caused a significant loss in the amount of data received from crashes.
Normally, when the Facebook app unexpectedly crashes on a user's device, as described in Section~\ref{subsec:fb-workflow}, telemetry about the device and application are logged and sent to Facebook's servers.
In this particular case, an on-call engineer noticed that in the 24 hours after the latest app update, the number of unexpected crashes in the Facebook app was up over 3 times compared to the previous version.
The engineer queried \name{} with crash reports from after this recent app update, and compared it to crashes from the previous app version.
Specifically, the engineer queried for 50,000 samples from each group of crashes, and mined patterns with a minimum support threshold of 0.02. In this case, metadata such as the OS version and app build version were concatenated to all traces.

\name{} identified several important insights: the new crashes, compared to others, were strongly correlated with older operating systems and empty data fields in many of the usually logged data fields.
The combined query for data and Minesweeper's algorithm took less than 2 minutes.
From there, the uptick in crashes was localized to a recent code push that had changed the JSON encoding system used when sending information about the app state, which had caused the loss of the data fields, and the incorrect code was quickly fixed.

Another instance where \name{} proved to be useful is when debugging low-signal errors; 
user-reported bugs fall under this category.
An on-call engineer decided to explore one particular cluster of bug reports (see Section~\ref{subsec:fb-workflow}). They collected 1500 traces from that cluster as the test group 
and a roughly equal number of traces from other clusters for the control group.
They ran \name{} with a minimum support threshold of 0\% (i.e., 1 trace), 
which would not be feasible with much larger input groups because of runtime constraints.

\name{} surfaced resulting patterns, which indicated that all of the bug reports were co-occurring distinctively with one particular type of server error related to permissions.
This permissions error had been repeatedly hit several times before the bug reports were submitted.
Upon examination into the bug reports associated with the cluster, many of the reports were indicating that they could not delete their pending Facebook group posts under some conditions. 
Because the symptom (not being able to delete posts) and the root-cause (a permissions error) were found, the on-call engineer was able to triage the now well-documented error to an engineer working on the affected parts of the app.
This engineer was then able to reproduce the issue with the information provided, and a fix to the incorrect permissions was quickly pushed. 

In both cases, \name{} simplified and aggregated the information to the on-call engineers so that they were able to root cause the issues quickly.
Specifically, the first error could have looked like an error with rolling out the new app version, but \name{} patterns instead pointed towards an anomaly in logging, saving valuable time by ruling out potential sources for error.
In the second case, \name{} correlated the bug to a particular error, which produced the signal needed for on-call engineers to root cause the problem.
Importantly, \name{} runs like these also happen fairly quickly, on the order of minutes at most, so that engineers working on time-sensitive tasks get results promptly, and less users are affected by potentially wide-scale disruptions.

\subsection{Linking regressions based on root cause}

\begin{figure}
\centering
\scalebox{0.8}{
\begin{tikzpicture}[
squarednodered/.style={rectangle, draw=red!60, fill=red!5, very thick, minimum size=5mm, align=center},
squarednodeblue/.style={rectangle, draw=blue!60, fill=blue!5, very thick, minimum size=5mm, align=center},
]
\node(plot){\includegraphics[trim=65 65 65 100, clip, width=\columnwidth]{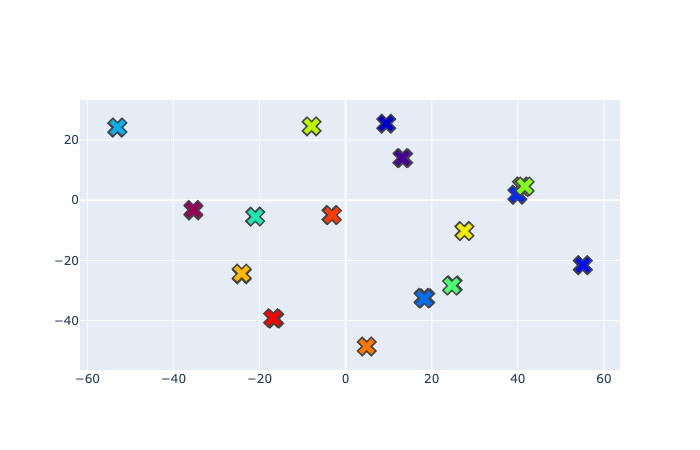}};
\node(a) at (-1.2,-1.2) [draw, red,line width=2pt,circle, minimum width=20pt, minimum height=20pt]{};
\node(b) at (0.55,1.8) [draw, blue,line width=2pt,circle, minimum width=20pt, minimum height=20pt]{};
\node[squarednodered](a1)[below=0.1cm of plot]{\footnotesize (a) ArrayIndexOutOfBounds, (b) IndexOutOfBounds on story media viewer};
\node[squarednodeblue](b1)[below=0.1cm of a1]{\quad \quad \footnotesize NPE on (a) FullScreenVideoPlayer, (b) FullScreenSeekBarPlugin \quad \; };
\end{tikzpicture}
}
\caption{Pattern-based representation of regressions in Facebook Android.}
\figlabel{vector-space}
\end{figure}

In our final experiment, we evaluate the accuracy of \name{} in linking together different regressions based on their root cause.
We use the \name{} pattern-based vector space model from Section~\ref{subsec:regressions} to encode regressions in the Facebook Android app in a time period of 3 months.
We then clustered the regressions within a cosine distance of 0.1 to each other.
A sample of the resulting clusters is shown in \figref{vector-space}, with two clusters 
highlighted as an example;
the points in the plot are too close, and so we highlight the clusters.
We found the clusters to contain regressions that intuitively share the same root cause, such as closely related exceptions on the same surface (red cluster), or the same exception on closely related surfaces (blue cluster).

To validate this quantitatively, 
we used data from tasks created for developers to track the fixing of these regressions.
Specifically, developers would attach code commits to these tasks and mark them closed once they put out a fix.
We gathered this set of code commits and made the following assessment: if two tasks were closed with the same commit, then they corresponded to the same, or very similar, regression.
Based on this metric, out of 34 pairs of regressions that were linked together in the clusters, 29 (85\%) were found to have the same commit as the fix, and 32 (94\%) were fixed by commits touching the same file(s).
This shows that \name{} patterns are highly accurate in identifying the root cause of regressions, and linking them based on it.
This also saves valuable time for engineers in debugging regressions that share the same root cause.

\subsection{Practical Application and Limitations}

While Minesweeper has been successful across many internal debugging workflows, 
we found that the quality of its results can be sensitive to various factors, outlined next.

First, Minesweeper expects aggregated and sanitized data as input. 
For example, if developers run Minesweeper on two groups of 100 traces each, 
and each input trace included exactly one of the hundreds of Facebook application components
visited by a user, we would expect poor results, for the same reason we preprocess numeric data (\ref{subsec:numeric}). In such cases, we instead aggregate the application components into 
larger related groups of components, such as news-feed, timeline, etc. Thanks to Facebook's mature logging infrastructure, we very rarely find the lack of aggregated or sanitized data to be a barrier to applying Minesweeper.

Second, Minesweeper's performance is adversely affected in case of poort signal.
 This can happen when the input data is noisy and patterns of interest are 
 not statistically significant from noisy input samples.
 A salient example is if, say in a group of 200 control and 200 regressing test traces, 
 5 traces are from Android devices and all have a specific crash, 
 whereas 195 traces are from iOS devices and only a few have the same crash. 
 In such a case, Minesweeper tends to miss the android crashes. 
 Additional domain knowledge from engineers, 
 for instance, restricting the test and control groups
 to Android devices only, can help navigate issues with poor signal.

Third, different parameter choices can, in some use cases, significantly affect the quality of results. Intuitively, Minesweeper is most powerful when it has the largest search space possible for every problem, which is accomplished by setting a low minimum support parameter, high maximum pattern length, and supplying as many input traces as possible. 
As expected, an obvious trade-off here is the increased running time for Minesweeper.

%% file: related.tex

\section{Related Work}
\label{sec:related}


Several previous works in the area of debugging and fault localization have used the notion of contrasting passing and failing executions.

The work of Liblit et al.~\cite{Liblit2005} 
identifies the root cause of bugs in a program from a
statistical analysis of its execution traces.  In Liblit's work,
the idea is to define Boolean predicates on program variables, branch outcomes, and function return values, and to evaluate them during program execution.
Then, given two groups of program runs -- failing and non-failing -- and their predicate evaluations, a statistical method is used to select predicates that are most likely explanations of the root cause of the failures.  In our work, likewise, we have to find statistically important patterns from potentially a large number of event patterns arising
from failing and passing traces.  

However, beyond this superficial similarity, the details differ substantially.  In Liblit's work, a large number of predicates might have equally high correlation with failing runs, but they may not be equally powerful in explaining the root cause, because some of these
predicates may be correlated with failures regardless of their value.  They use a notion of ``context'' to figure the background likelihood of a predicate being \textit{observed}, and discount for that.  In our setting, the concern comes up in a different way with redundant patterns, and we carry out the mitigation as explained in \ref{subsec:redundancy}.

Furthermore,  Liblit's work does not deal
with the time dimension, whereas the patterns on events that we seek are exactly temporal. 
On the other hand, since app telemetry has far fewer events than low-level
predicates as in Liblit's work, management of the proliferation of patterns is easier.

In addition to the work of Liblit et al., related work can be grouped into four ancestries.

First, there is a rich line of work on spectrum-based localization methods~\cite{DBLP:journals/tse/WongGLAW16}.  Among the best-known techniques in this category is Jone's et al's \emph{Tarantula} model~\cite{DBLP:conf/icse/JonesHS02}.  This technique creates a matrix whose rows denote different program locations touched during an execution---one could think of these also as control predicates---and the columns denote different executions.  Some of these executions are passing executions, and others are failing executions.  The goal of the technique then is to assign a \emph{suspicious-ness} score to each statement, based on its correlation with failure.   Jones et al. evaluated several ways to use this execution spectra to carry out fault localization.  More recently, however, Parnin and Orso~\cite{Parnin2011} showed that developers do not find such fault localization techniques useful in practice, especially for large programs.
\name{}'s view of predicates and traces is conceptually similar to spectrum-based techniques, but the idea of finding the fault is rooted in different statistical principles.

Second, there is work that applies statistical notion of a \emph{contrast set} on clusters of failing executions~\cite{DBLP:conf/sigsoft/CastelluccioSVP17,qian2020ccsm}.  Here, the idea is to use a initial clustering of failing executions based on some heuristics, but then use contrast sets to find what stands out in each cluster, compared to the normal statistics.  For instance, it could be the case that a certain cluster of crashes for a mobile app shows an anomalously high percentage of users using brand X device, compared to the overall percentage of users using brand X device.  Contrast set mining has limited power compared to \name{}, because it does not have any representation for temporal events.
Another related work, by Lin et al.~\cite{lin2020dataAI}, uses frequent itemset mining to find the subset of columns/features in a log, which all occur in multiple rows (which is the support of this item set) and are correlated with failures.  Their focus is on scalability and interpretability.  Again, Lin et al's work does not deal with finding temporal patterns.
(Note that \name{} is based on sequential pattern mining, rather than frequent itemset mining.)

Third, a vast literature on classifier learning---for example, decision-tree learning~\cite{Quinlan}---could in principle be trained to discriminate between passing and failing traces.  If successful, then may be techniques from model interpretability~\cite{ribeiro2016} can be used to further pin point which ``features'' of the inputs were more salient for failures.  However, this approach requires careful featurization of the input traces, which is essentially one of the key contributions of Minesweeper.

Fourth, \textsc{PrefixScan} and related data-mining algorithms have been used in a number of different applications; see Gupta~\cite{gupta2012applications} for a survey.  Minesweeper build on these algorithms, but to our knowledge, prior work has not leveraged these algorithms to troubleshoot software bugs at scale.

%% file: conclusions.tex

\section*{Conclusion}
We have proposed Minesweeper as a scalable tool for RCA. Minesweeper combines sequential pattern mining algorithms from the data mining literature and simple notions of statistical measures to isolate defining sequences that separate one group of traces from another. This tool allows engineers to quickly glean useful information from thousands of samples of telemetry in the order of minutes, which heavily speeds up regression debugging workflows and saves many end users from experiencing errors while using the Facebook app.

We discussed the binning of continuous data and the elimination of redundant output patterns, two important steps that make Minesweeper more flexible and useful to its users, as well as general use cases and parameter settings that allow for the best results. We discussed the performance of Minesweeper and showed that it performs well on both small and large datasets, on both fronts --- speed and quality of patterns.  
Finally, we discussed how Minesweeper helped root cause two specific bugs in the Facebook app.


